# Network Optimization Aspects of Autonomous Vehicles: Challenges and Future Directions

Rudolf Krecht, Tamás Budai, Ernő Horváth, Ákos Kovács, Norbert Markó, Miklós Unger

***Abstract*—Global megatrends, such as urbanization, population growth, emerging network solutions are accelerating the development of the Connected and Autonomous Vehicles (CAVs) industry. There are many truths, some misconceptions and to be honest, even some hype about CAVs in the public's opinion. The main objective of the current paper is to provide a comprehensive review, eliminate misconceptions and outline the future of the network optimization aspects of autonomous vehicles, by presenting various multidisciplinary methods such as cooperative perception. Given our extensive experience with CAVs, we are aiming to share some of the insights and knowledge we have gained, along with relevant use-cases and experiment results.**

## I. INTRODUCTION

COOPERATIVE perception has the potential to significantly improve the performance of autonomous vehicles in numerous ways. Firstly, by sharing information about the environment, autonomous vehicles can have a more complete and accurate understanding of the surroundings. Secondly, cooperative perception can help to reduce the individual sensor limitations, such as occlusion or noise, by fusing information from multiple sources. Novel methods, such as deep learning have been proposed with the aim of improving cooperative perception, but there are still numerous open questions on this field [1].

Network optimization is a critical aspect at the operation of CAVs, as it involves the efficient use of communication and sensor resources to support the safe and effective operation of these vehicles. This paper is aiming to discuss the challenges and future directions in the field of network optimization for CAVs, focusing on the application of communication and sensor networks to support cooperative perception and collaborative operation. Network optimization encompasses various techniques, approaches, and optimal practices to supervise, handle, and improve network performance. Traditional issues of network optimization incorporate reliability, bandwidth, utilization, and performance. The extended understanding also incorporates time synchronization, data representation and issues related to heterogenous architecture. The current paper addresses both traditional and extended network optimization approaches.

## II. OVERVIEW OF RECENT ADVANCES

Recent research advances regarding CAVs have shown that the known drawbacks of these vehicles can be significantly decreased using solutions from different related areas. The outstanding attention received by domains connected with autonomous vehicles has certainly helped the evolvement of CAVs as well.

### *A. Autonomous Vehicles*

A large part of the scientific community working in the field of autonomous vehicles agrees that these vehicles generate in the order of 10 GB data every minute. This amount of data requires fast storage and transfer. In the current data-driven era, large-scale datasets encourage research and development in this field. There are many available datasets [2], e.g., the KITTI dataset, Oxford RobotCar, Waymo Open Dataset, Argoverse, Boreas Dataset, USyd Dataset and so on [3]. These are

Manuscript received January 2023. The research was supported by the European Union within the framework of the National Laboratory for Autonomous Systems. (RRF-2.3.1-21-2022-00002). (*Corresponding author: Rudolf Krecht)*

Rudolf Krecht is with the Department of Automation and Mechatronics, Széchenyi István University, Győr, 9026 Hungary (e-mail: krecht.rudolf@ga.sze.hu)

Tamás Budai and Ákos Kovács are with the Department of Telecommunications, Széchenyi István University, Győr, 9026 Hungary (e-mail: budai.tamas@ga.sze.hu; kovacs.akos@ga.sze.hu)

Ernő Horváth, Norbert Markó and Miklós Unger are with the Vehicle Industry Development Center, Széchenyi István University, Győr, 9026 Hungary (e-mail: herno@ga.sze.hu; marko.norbert@ga.sze.hu; unger.miklos@ga.sze.hu)

solely singe vehicle-based datasets, but there are some multi-vehicle measurements as well, in case of which the results are usually simulated, like in case of the Cooperative Driving Dataset [4]. Recent trends include advanced sensors (e.g., solid-state LiDAR), purpose-built vehicles (like Zoox and Cruise, Fig. 1) and advanced teleoperation that ensures reliable human intervention. In the 2010's, at the early days of autonomous vehicles, spinning 16-64 channel LiDARs meant the state-of-the-art distance sensing solution. In the last 10 years, the number of patents regarding LiDAR sensors has tripled. During the same period, LiDAR technology showed a substantial price drop as well. Although sales volumes have increased, this price drop was not the result of mass-production. The more likely reason is that the technology itself evolved, the production technology changed. The solid-state LiDAR is one of the most prominent results of this technological shift. A solid-state LiDAR has no moving parts, but a fixed field of view. The introduction of solid-state LiDARs has significantly increased the lifetime of these sensors as well. Since there are no moving parts, a solid-state LiDAR is expected to run 100,000 hours. In contrast, a spinning LiDAR may typically run between 1,000 to 2,000 hours.

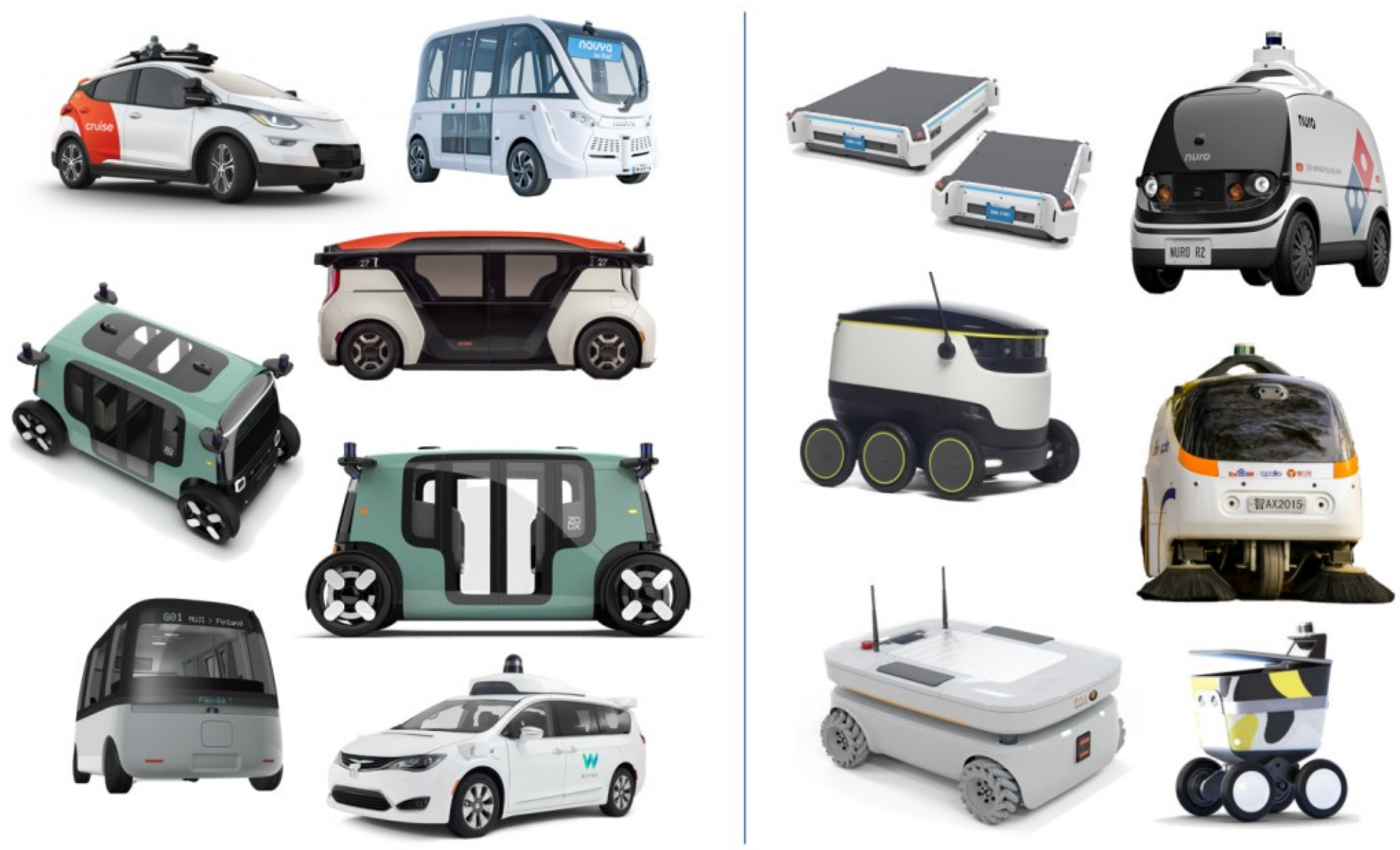

**Fig. 1.** Passenger (left) and Non-passenger CAVs (right). The companies are Cruise, Zoox, Waymo, Sensible4, Navya, Fetch, Nuro, Idriveplus, Tiago, Postmate, Starship

Purpose-built vehicles are one of the recent trends of autonomous vehicle development. This means that there are vehicles which were designed to be autonomous, most notably they do not even have steering wheels, as visible in Fig. 1. This also allows symmetric design in some cases (e.g., Zoox, Cruise). Symmetric design means that there is no front part, the vehicle can drive both ways with two-axle steering. This also means that less different parts need to be designed or manufactured.

*B. Machine Learning for Connected Vehicles*

Technological advances in the last decade have opened new perspectives regarding possibilities in the field of autonomous driving. Dramatic increases in computational performance along with some long-existing ideas applied in novel ways spawned a surge of new research in the field of artificial intelligence, and it was a natural consequence that existing problems for connected vehicles also benefitted from this artificial intelligence boom.

A popular categorization divides the subtasks of autonomous driving as perception, localization, planning and control. All of these subcategories might benefit from implementations based on artificial intelligence, but the most fundamental and logical connections might be observable in case of unified perception.

An interesting challenge that has been approached from different perspectives is common scene completion. Apart from a meticulously orchestrated setting for a cooperative perception experiment, the sensors of various vehicles populating the roads can vary tremendously even when talking about the same sensor technology. These distinctions range from fundamental differences, like non-corresponding resolution or aspect ratio values in case of camera sensors, to completely different sensor constellations that are even difficult to compare, let alone matching their output data in real time. The sensor setup can certainly be standardized among the vehicles of large automakers, but the goal of any scientific research towards cooperative perception should be a car brand agnostic solution, therefore this approach has been promoted.

To understand how artificial intelligence can contribute to the achievement of this goal, the related concepts have to be clarified. At its core, robotic navigation requires geometric and semantic environmental perception to interact with and navigate in the real world successfully. This leads us to the greater idea of 3D scene understanding, the ultimate reason why vehicles are being equipped with sensors. In case of a single robot, this task is much simpler due to known rigid transformations between sensor coordinate frames and due to planned sensor setups. Multiple units add several layers of complexity to this problem, since base frames are moving relative to each other and information received from sensors mounted on other vehicles might not always be useful depending on distance, angle, information type etc. Humans are more capable to estimate geometry and semantics even in case of largely occluded regions. This is achieved by prior knowledge leveraging, and this is where artificial intelligence, more specifically semantic scene completion comes into the picture. Semantic scene completion (SSC) is a type of reconstruction where semantics are also inferred along with geometry, and this is done for both seen and unseen regions [5]. Since deep learning has the capacity to store learned priors, they prove to be incredibly useful in replicating the mentioned human capability to infer occluded areas. With these considerations, it will be introduced how artificial intelligence can help cooperative perception and give solutions to the problems that arise during network usage.

First, SSC would be tremendously helpful in the utilization of incomplete information, since reconstructing additional parts of the scene with large chunks of missing data would often serve as noise at best and contradictory information at worst. Using neural model priors could make incomplete data useful. In addition to that, deep learning can help matching the coordinate frames, which seems to be a simple task, but matching moving frames can become really complicated in real time. Lastly, predictions from other vehicles can also be utilized in addition to raw sensory data.

Even with the rapid development of networking solutions, it is important to consider bandwidth limitations. When using SSC, there are several different data representations and input encodings. All these solutions have different trade-offs regarding information density and network usage. A good example of this would be the tradeoff between representing the 3D space with a point cloud versus an occupancy grid. The occupancy grid (voxel grid) has significant memory needs which stems from the fact that both occupied and free space needs to be represented. The description of the relationship of grid cells arises organically from the nature of this grid ordering. Point clouds on the other hand are sets of points lying on the surface which do not represent free space or geometrical structures, but they are more memory efficient compared to voxel grids which makes it easier to broadcast them through networks with limited bandwidths. A more recent memory efficient data representation is presented in [6], where point clouds are transformed to convolutional features before matching, thereby decreasing the memory and bandwidth requirements dramatically.

In addition to choosing a suitable representation for our needs, applying a sparse data structure, like sparse matrix or tensor, can significantly reduce memory, bandwidth, and computational resource usage [7].

### *C. Connectivity Frameworks for Connected and Autonomous Vehicles*

In the case of autonomous and interconnected vehicles, a connectivity framework can be defined as a standard communication protocol that allows the seamless integration of new elements of infrastructure and vehicles. The spreading of connected vehicles induced the necessity of standard connectivity frameworks between the various interconnected components. Since there is no generally accepted, standard connectivity framework, there are many open research topics on this area.

Most of these approaches use Data Distribution Service (DDS) as a starting point. DDS is a specification for a publish-subscribe type data distribution system. The aim of this system is to create a platform-independent model for data handling. These data handling solutions apply *Publishers* and *Subscribers* for data handling. A Publisher is responsible for issuing the data. A Publisher is equally responsible for issuing correct data values at appropriate time intervals. A Subscriber receives the initially published data and makes it available in the format requested by the user. The Publisher and the Subscriber can be implemented on different, even remote systems.

A good example of the application of DDS is Robot Operating System 2 (ROS 2). By default, ROS 2 uses DDS as its middleware. ROS 2 is a widely accepted and applied software for robot development, and it has become the de-facto standard for autonomous vehicle development as well. It consists of a set of open-source software libraries that offers a wide variety of tools and solutions for frequent problems and tasks in the field of self-driving vehicles. ROS system can use TCP and UDP transfer as well to share data between ROS nodes, but to optimize network usage, UDP has become a widely used protocol among ROS users.

Applications based on DDS show an emerging tendency, and even though, as previously mentioned, it is not an official standard, but it has been accepted as a general solution in case of research and development in the field of connected self-driving vehicles [8].

## III. Open Research Challenges

Since there availability of possible solutions regarding communication and environmental perception is increasing, the number of open research questions is showing an emerging tendency as well. One major challenge in network optimization for CAVs is the need to support a wide range of applications and scenarios with varying requirements in terms of communication and sensor performance. For example, some CAV applications may require high-bandwidth and low-latency communication for *real-time decision making*, while streamed multimedia requires only high-bandwidth, and finally, some additional service data may not be sensitive for data rates or latency either. Similarly, some CAV applications may require high-resolution and high-precision sensor data, while others may be able to operate with lower-resolution and lower-precision sensors. Another key challenge in network optimization for CAVs is the need to support heterogenous devices and applications in a highly dynamic environment.

### *A. Parallel Usage of V2X and 5G Technologies*

V2X (Vehicle-to-Everything) is a term that refers to the various communication technologies that enable vehicles to communicate with each other and with their surroundings [9]. This includes technologies such as Vehicle-to-Vehicle (V2V), Vehicle-to-Infrastructure (V2I), and Vehicle-to-Pedestrian (V2P) communication. DSRC (Dedicated Short-Range Communication) is a type of V2X (Vehicle-to-Everything) technology that allows vehicles to communicate with each other and with their surroundings using short-range wireless communication. DSRC uses a dedicated portion of the radio spectrum (5.9 GHz band) for communication, and it was designed mainly for safety-critical applications such as collision avoidance and lane change assistance.

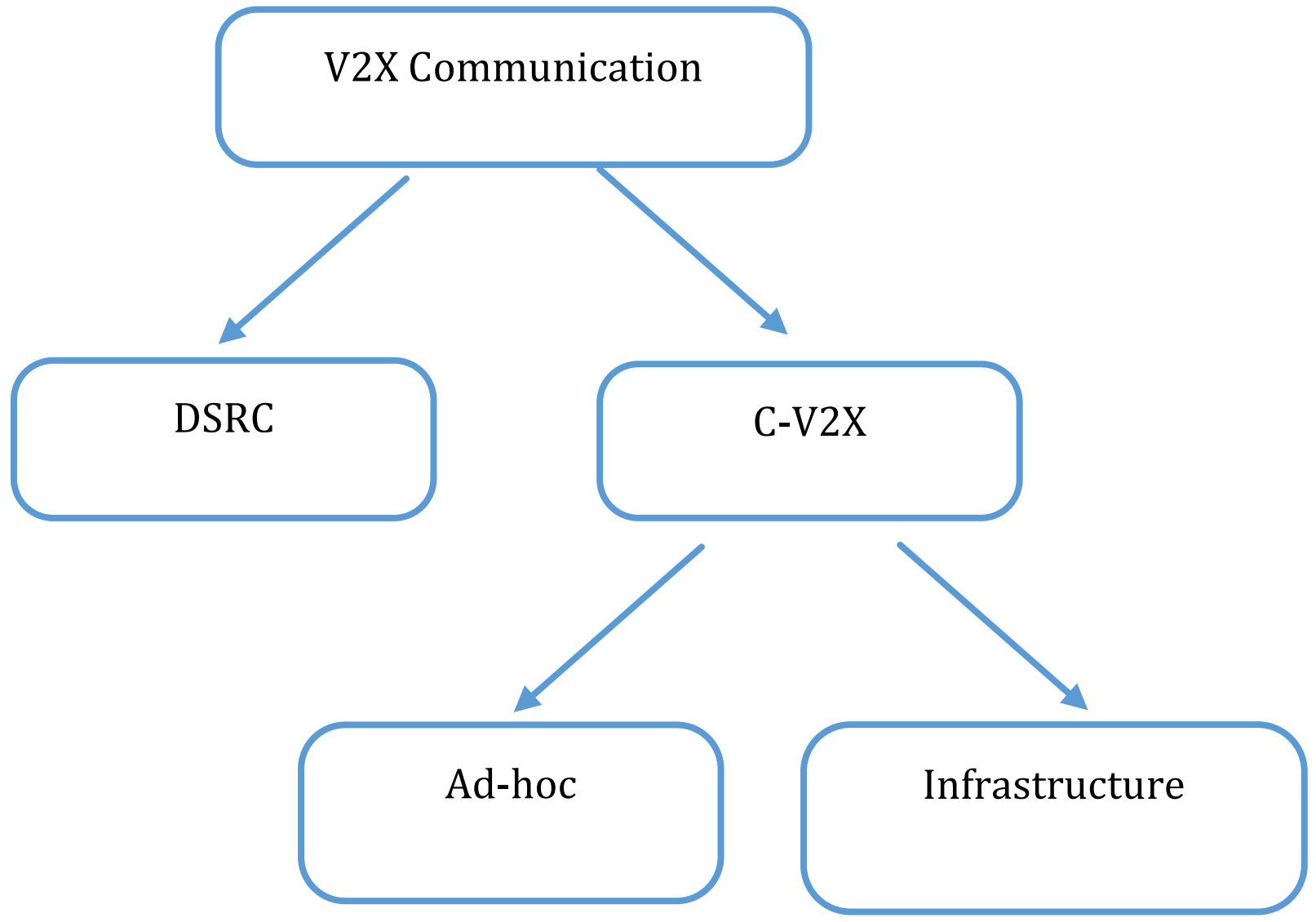

**Fig. 2.** An overview of communication technologies

DSRC technology has been deployed in some regions for use in transportation systems and it is seen as a promising technology for improving road safety and efficiency [10]. However, DSRC has a limited range and may not be suitable for all V2X applications.

C-V2X (Cellular Vehicle-to-Everything) is a specific type of V2X technology that uses cellular networks (such as 5G) to enable communication between vehicles, infrastructure, and other devices. C-V2X is often seen as a more reliable and efficient alternative to traditional V2X technologies, as it can take advantage of the robustness and wide coverage of cellular networks.

In summary, V2X refers to any technology that allows vehicles to communicate with their surroundings, while C-V2X is a specific implementation of V2X the technology that uses cellular networks to enable communication [11]. Even though Connected Vehicles generate a vast amount of data, V2X technologies specify very strict formats which are not optimized for sharing data types often applied in case of environment sensing, such as raw LiDAR point clouds. Using 5G to transfer the aforementioned group of data types is necessary to handle difficult traffic situations.

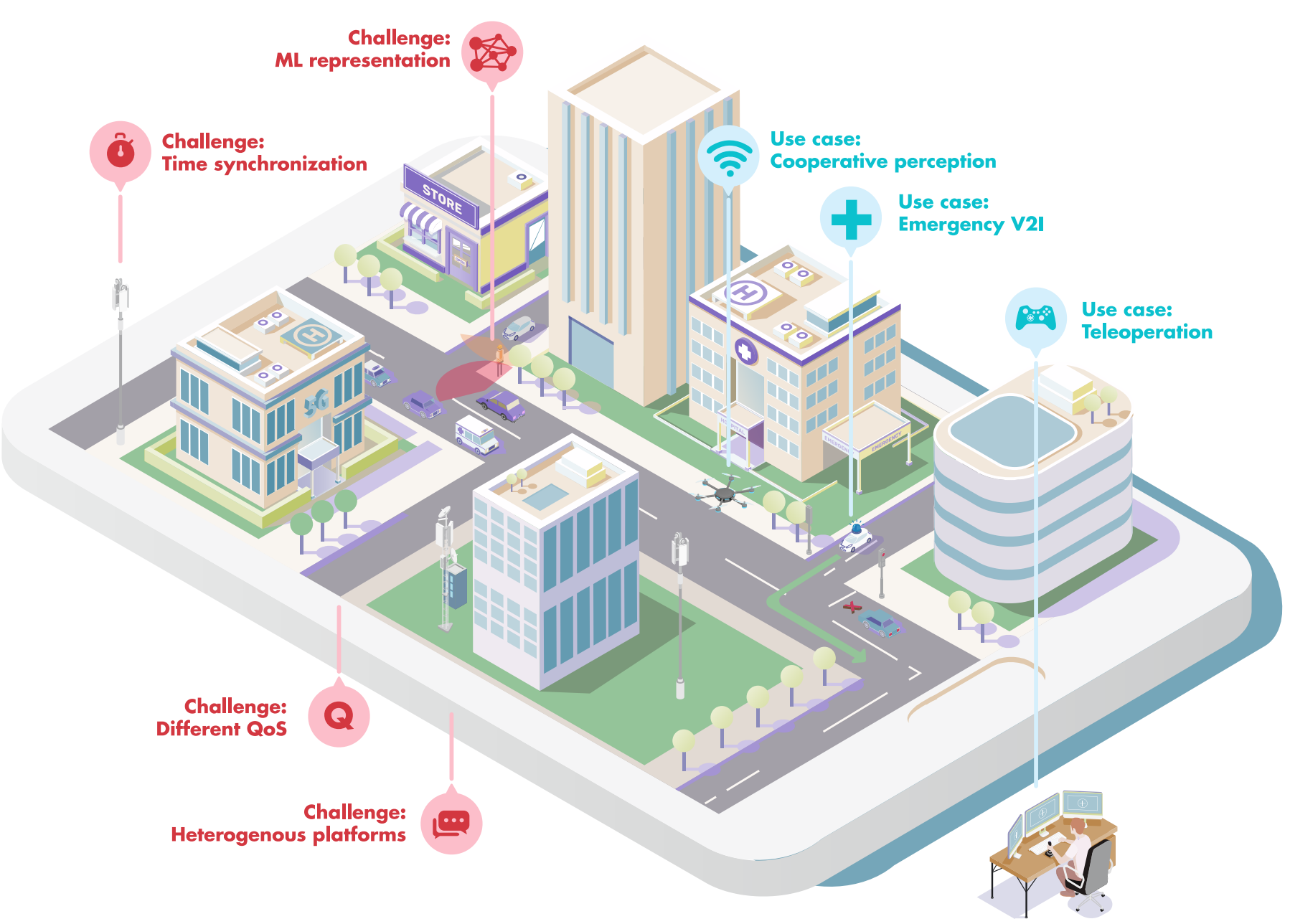

**Fig. 3.** Illustration of CAVs' network challenges together with the described use-cases

*C. Technical Issues of CAV Development*

Time synchronization is an important aspect of communication systems, as it allows different devices to coordinate their actions and exchange information in a coordinated manner. In 5G (Fifth Generation) cellular networks, time synchronization is achieved using a combination of hardware and software solutions.

One potential issue with time synchronization in 5G networks is that the high data rates and low latency requirements of 5G may make it challenging to maintain accurate time synchronization. This is because the timing of communication events in 5G networks is often determined by the transmission and reception of high-speed data, which can be affected by various factors such as noise and interference.

To address this issue, 5G networks may use advanced timing synchronization techniques such as GPS synchronization and precise time protocol (PTP) to maintain accurate time synchronization. Additionally, 5G networks may use multiple frequency bands and multiple antennas to improve the reliability of time synchronization [12].

Overall, maintaining accurate time synchronization in 5G networks is an important challenge, but it can be addressed using advanced synchronization techniques and the deployment of robust hardware and software solutions.

Using different network technologies with different QoS (Quality of Service) parameters is challenging, as it is not straightforward to select the best data path between two endpoints in any given point in time because many factors must be considered based on the nature of the data to be transferred [13]. Different information types may have different requirements regarding transfer speed, latency, reliability, and robustness. A promising way to answer this is to introduce a higher level of abstraction on the network level, that covers advantages and disadvantages of the different networks such as MPTCP. Multipath TCP (MPTCP) is a transport-layer protocol that allows a single TCP connection to use multiple paths for data transmission. This can provide several advantages in certain communication scenarios [14]. Some of the main advantages of MPTCP are:

- Improved reliability: By using multiple paths for data transmission, MPTCP can increase the reliability of communication by providing redundant paths for data transmission. If one path fails, the data can be transmitted over the remaining paths.
- Increased bandwidth: MPTCP can increase the effective bandwidth of a connection by allowing data to be transmitted over multiple paths simultaneously. This can be particularly useful in scenarios where multiple paths with different bandwidths are available.
- Improved network utilization: MPTCP can improve the utilization of network resources by allowing multiple paths to be used simultaneously, rather than requiring separate connections for each path.
- Better performance: MPTCP can improve the performance of communication by allowing data to be transmitted over the path with the lowest latency or the least congestion.
- With MPTCP different paths can be weighted according to the bandwidth and reliability of the path so different paths can be optimized for different data transfer tasks.

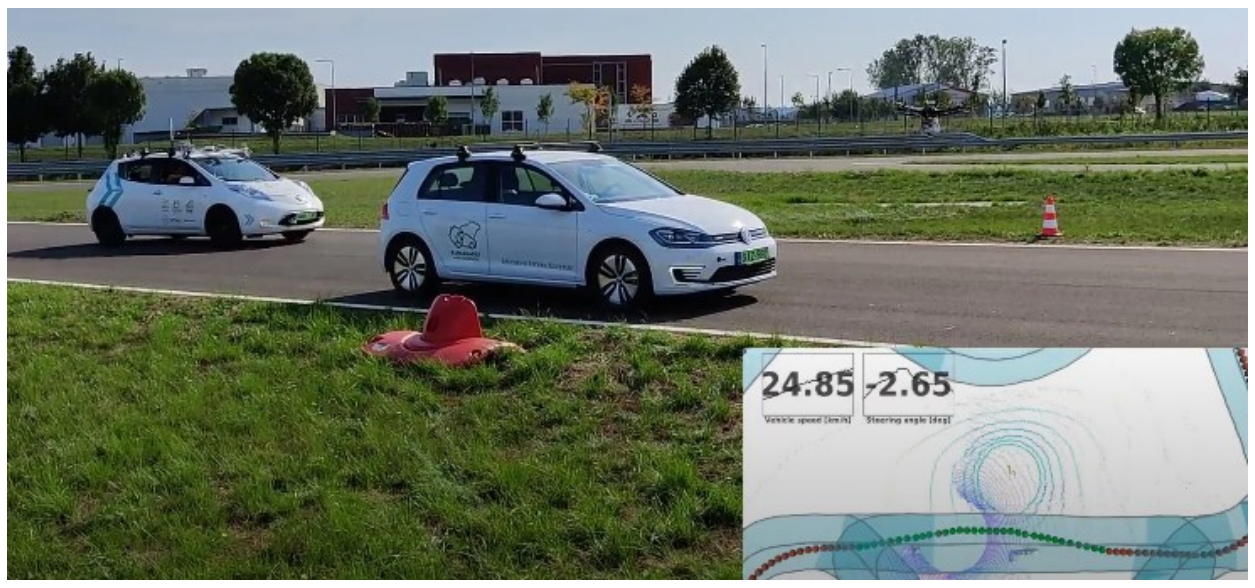

**Fig. 4.** A CAV avoiding a halted vehicle with the assistance of a connected UAV

### *D. Automated Ad-Hoc Networks and Data Relevance*

During our work, the communication paths between all participants of the introduced scenarios (e.g., CAV, drone, traffic lights) were established by configuring network equipment manually. As the scenario was predefined, information about the number of participants and the required communication paths has been available in advance, therefore it was straightforward to come up with a network topology that fulfills all requirements and is also optimal in terms of utilization of different communication technologies.

When V2X systems will be deployed and will be utilized to aid live traffic on the roads, this information will not be available, since the actors of the emerging traffic situation and the necessary connection paths are not known in advance. Network topology and these paths must be established in an automated way in an often very short time frame as the traffic situation in question unfolds and must be torn down as soon as the situation is resolved [15].

Another important aspect that is closely linked to this previous one is the relevance of different parts of the transmitted information. In most cases, most of the data that must be exchanged between the participants of a given traffic situation is only relevant in the geological proximity of the situation itself. There is no need to transmit all this data to other endpoints, as it only leads to unnecessary network loads. However, some of the collected and exchanged data that is used to resolve a traffic situation may be relevant and useful for other purposes where the entities that wish to receive and process this data are either not in the physical proximity of the situation or it is impractical to tie the entity to a location given the distributed nature of it (e.g., cloud services). These entities can use the data during higher level decision making, such as road traffic engineering or route planning.

This poses two challenges; selecting the parts of the exchanged data that should be transferred to other entities and establishing an optimal communication path to transfer the data to these other entities over the different networks mentioned earlier.

## IV. Case Studies

Selected aspects of network optimization are shown through the presentation of case studies. These case studies present some examples managing the challenges described earlier. These issues incorporate network performance, time synchronization, data representation, and heterogenous architecture-related matters.

The selection of these case studies has been carried out with the aim to showcase the multitude of possible approaches for overall network performance enhancement.

### *A. Cooperative Perception between an Aerial Non-passenger and a Passenger CAV*

This case study shows the most typical practical issues of cooperative perception: data representation, time synchronization and heterogenous architecture. In the current case study, a passenger CAV is communicating with a connected Unmanned Aerial Vehicle (UAV). The use case was set up for obstacle detection, enabling the CAV to perceive occluded or distant obstacles. The aim of this case study was to demonstrate how UAVs could help CAVs to solve problematic, real-world traffic situations. In our experimental scenario, a static obstacle (e.g., a vehicle forced to stop by a technical defect) was placed in the path of a self-driving vehicle. Thus, the self-driving vehicle had to carry out an evasive maneuver. The UAV was flying in the proximity of the static obstacle, and it was able to communicate with the self-driving vehicle (Fig. 4). Both the UAV and the self-driving vehicle were equipped with an RTK-GPS receiver and a LiDAR sensor. In addition, the vehicle also used cameras. As a technical detail, the two vehicles communicated via ROS, only one *rosmaster* was running, which means that all UAV-related information was accessible by the self-driving car, and accordingly, all car-related data were accessible by the UAV. As the two entities of the communication were ROS nodes, UDP was selected for the underlying communication layer instead of TCP. The static obstacle was always in the range of the UAV, and the approaching vehicle was able to detect the obstacle with LiDAR-camera fusion. Despite the heterogenous perception, common interfaces and coordinates described in the ROS REP 105 guaranteed that the shared obstacles are understanded similarity. Based on GPS input, it was possible to publish the position of the obstacle as a *rostopic* in global coordinates, with synchronized timestamp which, was accessible by the passenger CAV as well. Thus, the CAV received information about an obstacle sooner than it could have been detected by its own on-board sensors. The significance of this is that the sooner the position information is received, the sooner the avoidance planning algorithm can calculate the avoidance path. In the current case data representation was chosen to be network effective, lacking cooperative representation. This means no pre-processed sensor data was shared, only the fully processed obstacles. From a network point of view, it is important to mention that the drone sends only one and indispensable topic to the self-driving vehicle. This topic contains the position of the detected obstacle. The limitation of this method is that all sensors have to be synchronized via ROS, otherwise there could be inaccuracies in the position data of the vehicle or the obstacle, leading to an unsuccessful evasive maneuver.

### *B. V2I Communication between Emergency Vehicle and Traffic Light*

Vehicle prioritization at intersections may not solely be a use case of CAVs, but also of connected vehicles. Although our use case incorporates an autonomous vehicle, a non-autonomous counterpart of the same scenario could be imagined. A V2I communication between a traffic light and an emergency vehicle has been set up. The aim was to demonstrate how V2X communication can increase traffic safety during the missions of emergency responder vehicles (e.g., police car, fire engine, ambulance).

In case of our solution, a communication tool was responsible for continuously transmitting a distinguishing signal during the emergency vehicle's mission. As the vehicle entered the range of the sensor mounted on the traffic light, the corresponding traffic light switched from red to green automatically, while the other lights switched to red to stop cross traffic. This experiment demonstrates that future autonomous vehicles will not require specific perception algorithms to handle traffic situations involving active emergency responder vehicles. Instead, these situations can be resolved by V2X communication, which allows the reduction of direct data transfer between vehicles.

### *C. 5G Teleoperation of a Passenger Vehicle*

5G potentially has the bandwidth, transmission time, jitter, and reliability to teleoperate a distant connected vehicle. Teleoperation can be either remote assistance (sometimes referred as teleguidance) or remote control. Remote assistance is an indirect control, where the operator gives high-level commands to the vehicle. An example for that is when a CAV fails to estimate the easing of a congestion or the removal of an obstacle. A human operator can give the high-level information whether a stopped car persists or goes away quickly. Remote control, on the other hand, gives direct commands (wheel movement, brake, or acceleration) to the CAV. Remote control can be extremely sensitive to network latency or bandwidth, while remote assistance has slightly looser requirements.

By analyzing our teleoperation use case, several issues of network optimization can be comprehended more easily. In specific situations in which a CAV cannot operate safely, a remote operator could take over the control or provide additional information for the vehicle. In case of our experiment, a remote driver not only is provided with real-time camera feed from the teleoperated vehicle, but also with the visualization of the raw LiDAR point cloud stream, current velocity, steering angle, and the current pose of the obstacles. During our demonstration, the vehicle was driven from roughly 150 km away. To ensure the best results, a 5G campus network was used for this scenario. This network allows communication between network entities (CAVs and remote operators) on layer 2 level, which eliminates the overhead present in case of different VPN solutions. In our case, teleoperation was achieved with an average of ~30 ms roundtrip transmission time (~120 ms including software-related and actuator-related latency) from Zalaegerszeg to Győr. This latency is satisfactory to teleoperate a vehicle at velocity levels common for urban areas.

All the mentioned network connections are shown in Fig. 5.

## V. Future Directions

The rapid development of V2X solutions in the past few years has resulted in numerous patents and standards and has reached a maturity level on which is necessary to build hardware and software solutions based on these standards. Based on these tendencies and our use cases, the future of CAVs have been analyzed by categorizing them as non-passenger and passenger vehicles.

### *A. C-V2X and Private 5G*

Nowadays one can find multiple vendors who provide V2X hardware and software solutions for car manufacturers (On-board units) and for public road operators (Roadside units). These modules are equipped with modern software-defined radios and chipsets to support future changes in the ever-evolving V2X standards.

Despite the wide availability of these devices, deployments are not yet widespread because the ecosystem is struggling with a mutual dependency problem; as long as there is no infrastructure alongside public roads, car manufacturers are not motivated to include On-board units in their vehicles, however as long as there are not enough vehicles equipped with V2X technology, public road operators are not motivated to invest in infrastructure development.

This situation together with the fact that CAV experiments are generally allowed on public roads for safety reasons, led us to the conclusion that all currently available communication technologies, such as 5G, should be employed to model the communication capabilities of upcoming V2X technologies to be able to experiment on real vehicles on closed roads.

One promising example of this is the combined usage of DSRC and private or campus 5G networks using network slicing to ensure the reliability and security of the communication.

### *B. Future Direction of CAVs*

It is important to distinguish non-passenger CAVs and robotaxi-like passenger CAVs (Fig. 1). Non-passenger automated CAVs are already being utilized and are expected to see widespread adoption. These CAVs will be used for inter-city logistics, urban "middle-mile" transportation, last-mile delivery robots, and off-road freight applications (e.g., airports, warehouses, and mines). These automated freight CAVs will address the shortage of truck drivers and increase transportation efficiency. Other examples of non-passenger CAVs are present in waste collection, snow removal, agriculture, grass cutting, etc. The implementation of these vehicles is strongly justified from an economical perspective, and safety concerns might be easier to address than in the case of self-driving passenger vehicles.

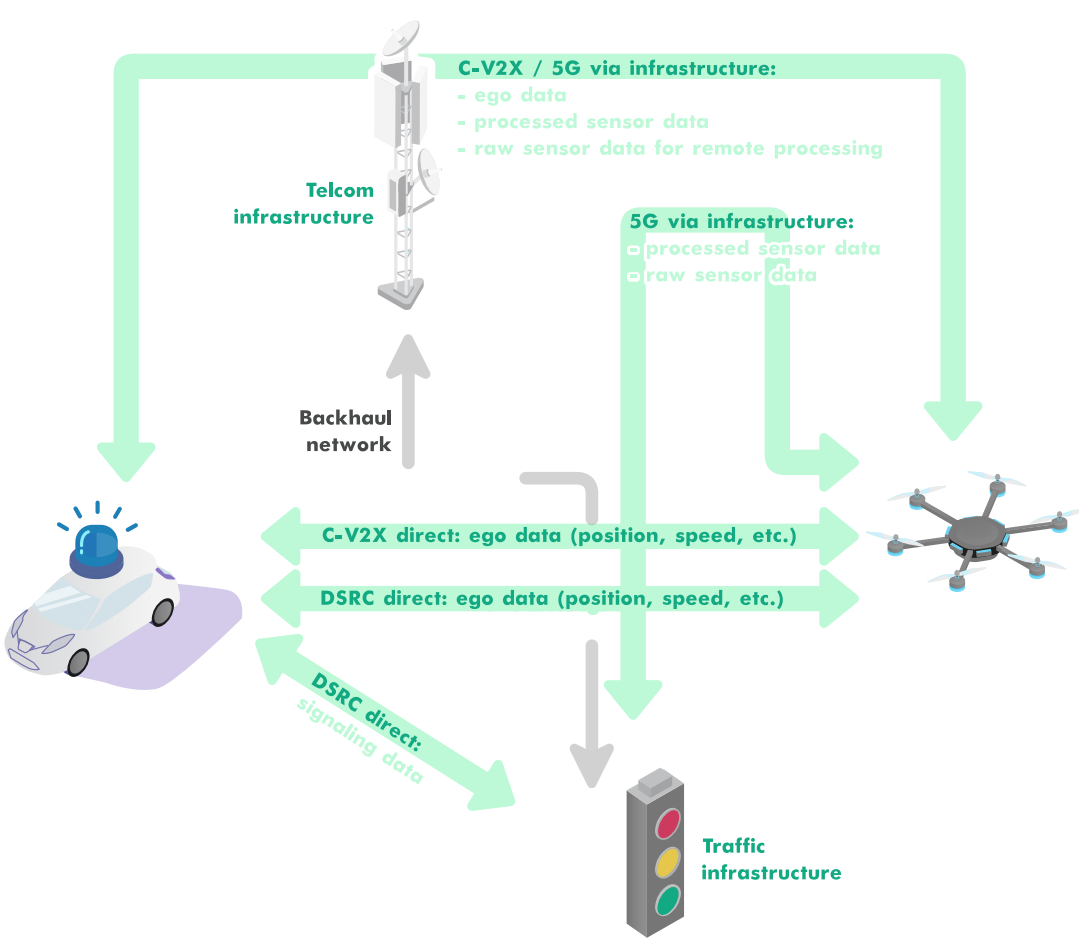

**Fig. 5.** High-level network architecture diagram

Many successful robotaxi-like passenger CAV services have been deployed in developed countries, with some of them having fleets of thousands of vehicles. This seems to be a high number, but compared to human-driven taxis, there is still a potential for growth. From a business point of view this solution is marginally or not even profitable, but in the long run it is definitely expected to be. This means that the widespread of non-passenger CAVs is expected in the close future, followed by passenger ones in the 2030s.

## VI. Conclusion

There are still many open questions in the world of connected autonomous vehicles. These questions range from the quality of communication, through transmission protocols, to the refinement of machine vision and sensing, and the further development of applications based on artificial intelligence. These open questions will continue to hinder the widespread adoption of connected autonomous cars for a long time to come, while at the same time motivating researchers to constantly push the boundaries of this field.